\documentclass[lettersize,journal]{IEEEtran}
\usepackage{amsmath,amsfonts}
\usepackage{algorithmic}
\usepackage{algorithm}
\usepackage{array}
\usepackage[caption=false,font=normalsize,labelfont=sf,textfont=sf]{subfig}
\usepackage{textcomp}
\usepackage{stfloats}
\usepackage{url}
\usepackage{verbatim}
\usepackage{graphicx}
\usepackage{cite}
\usepackage{paralist}
\usepackage{siunitx}
\usepackage{bbding}
\usepackage{makecell}
\usepackage{soul}
\usepackage{enumitem}
\usepackage{hyperref}

\usepackage{cleveref}
\newcommand{\chirev}[1]{{\color{black}{#1}}} 
\newcommand{\vis}[1]{{\color{black}{#1}}} 
\newcommand{\visrev}[1]{{\color{black}{#1}}} 
\newcommand{\tvcgminor}[1]{{\color{black}{#1}}} 
\newcommand{\tvcgminorr}[1]{{\color{black}{#1}}} 
\begin{document}

\title{Narrative Player: Reviving Data Narratives with Visuals}

\author{
Zekai Shao,
Leixian Shen, 
Haotian Li, 
Yi Shan, 
Huamin Qu,
Yun Wang, 
and Siming Chen

\thanks{
Z. Shao, Y. Shan, and S. Chen are with Fudan University.
S. Chen is also with Shanghai Key Laboratory of Data Science.
E-mail: zkshao23@m.fudan.edu.cn; \{yshan20, simingchen\}@fudan.edu.cn.

Y. Wang is with Microsoft.
E-mail: wangyun@microsoft.com. 

L. Shen, H. Li, and H. Qu are with The Hong Kong University of Science and Technology.
E-mail: \{lshenaj, haotian.li\}@connect.ust.hk; huamin@cse.ust.hk.

S. Chen and Y. Wang are the corresponding authors.

Work done during Z. Shao and L. Shen’s internship at Microsoft.
}}

\markboth{Journal of \LaTeX\ Class Files,~Vol.~14, No.~8, August~2021}%
{Shell \MakeLowercase{\textit{et al.}}: A Sample Article Using IEEEtran.cls for IEEE Journals}


\maketitle

 
\begin{figure*}[ht]
\centering
\includegraphics[width=\textwidth]{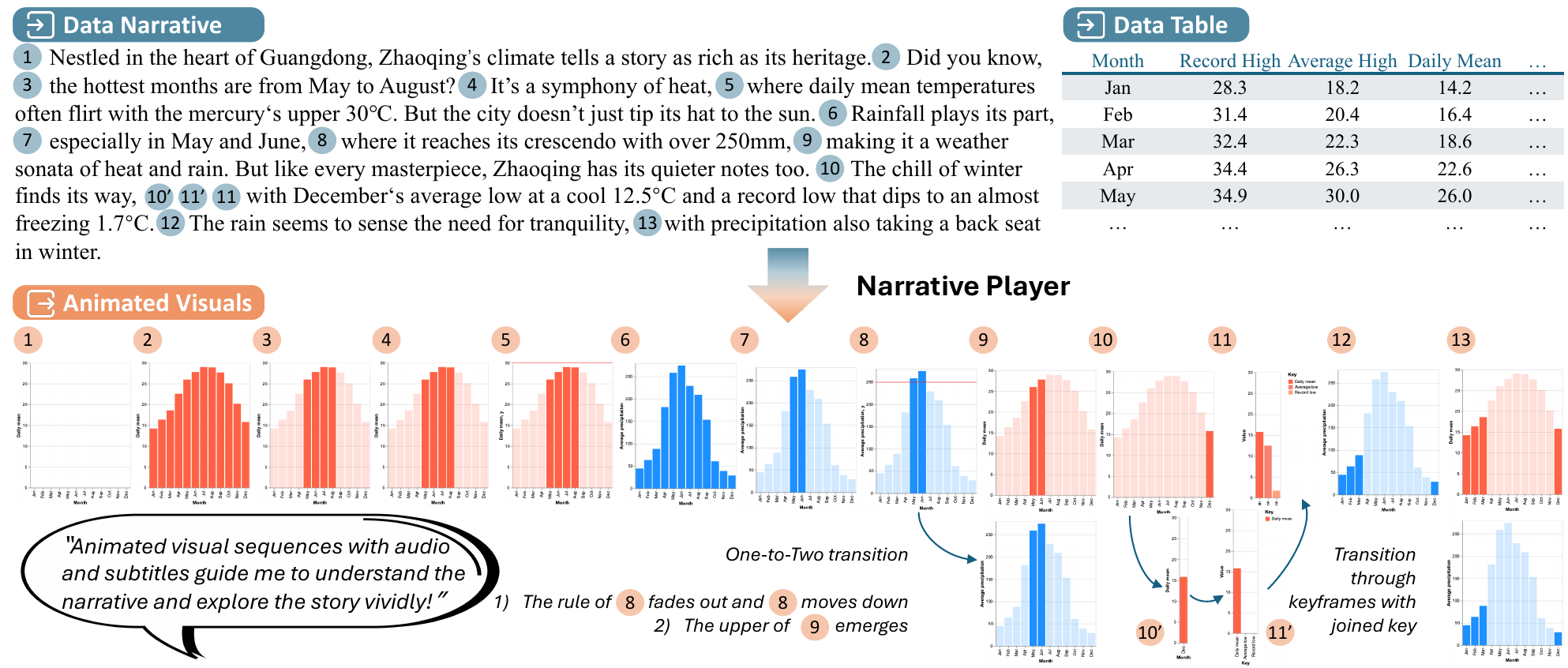}
 \caption{\tvcgminor{An animated visual sequence example with seamless transitions, audio and subtitles automatically generated by Narrative Player from data narrative and data table for engaging reading experience, where \raisebox{-0.12cm}{\includegraphics[height=0.4cm]{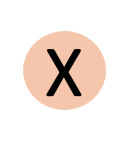}} means the visuals will transition to the next when the narration moves forward to the corresponding $x$-th segment.}}
  \label{fig:teaser}
\end{figure*}
 
\begin{abstract}
\vis{
Data-rich documents are commonly found across various fields such as business, finance, and science. 
However, a general limitation of these documents for reading is their reliance on text to convey data and facts. 
Visual representation of text aids in providing a satisfactory reading experience in comprehension and engagement. 
However, existing work emphasizes presenting the insights within phrases or sentences, rather than fully conveying data stories within the whole paragraphs and engaging readers.
To provide readers with satisfactory data stories, this paper} presents Narrative Player, a novel method that automatically revives data narratives with consistent and contextualized visuals.
Specifically, it accepts a \vis{paragraph} and corresponding data table as input and leverages LLMs to characterize the clauses and extract contextualized data facts. 
Subsequently, the facts are transformed into a coherent visualization sequence with a carefully designed optimization-based approach.
Animations are also assigned between adjacent visualizations to enable seamless transitions.
Finally, the visualization sequence, transition animations, and audio narration generated by text-to-speech technologies are rendered into a data video.
The evaluation results showed that the automatic-generated data videos were well-received by participants and experts \vis{for enhancing reading}.
\end{abstract}



\section{Introduction}
\vis{
\IEEEPARstart{D}{ata-rich} documents, scientific reports, and a wide range of reading materials encompassing fields such as business, finance, and science, are commonly encountered in daily life. These data narrations serve as a means for readers to comprehend the underlying data stories, thereby facilitating the knowledge acquisition and decision-making processes.

However, the conventional text-based format of data communication has several limitations when it comes to data comprehension. It can be difficult to interpret, fragmented, and prone to misinterpretation~\cite{Masson2023,macdonald1977numbers,tufte2001visual}. Moreover, lengthy text often leads to a monotonous reading experience, causing boredom and a lack of focus, which negatively impacts user engagement~\cite{gyselinck1999role}.
In contrast, visual representations of text, such as illustrations and visualizations, aid in constructing mental models of the information and enhance comprehension~\cite{glenberg1992comprehension, gyselinck1999role, duke2009effective}. 
Furthermore, visualizations have played a significant role in distant reading~\cite{moretti2013distant} by transforming text into an abstract view that highlights overview features. Dynamic visualizations and videos take this a step further, actively engaging readers and enhancing data communication through live storytelling~\cite{data_player,ying2023reviving}.
Considering these factors, we are motivated to generate live data stories from data-rich documents, aiming to enrich the reading experience and enhance user engagement \tvcgminor{in \tvcgminorr{terms of} understandability, memorability, focused attention, and enjoyment~\cite{ying2023reviving}.}

Given the potential benefits of integrating text with visuals, multiple studies have proposed solutions to enhance the reading experience of archival documents from diverse perspectives.
One line of research employs annotated visualizations primarily for navigation~\cite{Hullman2013, El-Dairi2019}, offering brief annotations to situate a document within a larger corpus instead of exploring the document's specific story or context.
This approach's superficial grasp of the context limits its ability to generate visuals that enrich the reading experience of the data story.
Others focus on reading within individual small data-rich subsets of documents, but ignore the broader data narratives ~\cite{elasticdocuments, Masson2023}.
For instance, Elastic Documents~\cite{elasticdocuments} establishes cross-references between tables and text through keyword matching and generates on-demand visualizations for user-selected sentences. 
However, these methods focus on enhancing comprehension via visualizing local data insights \tvcgminor{within phrases and sentences}, rather than the global text context \tvcgminor{within overall paragraphs}, and they do not emphasize reading enjoyment with vivid data stories driven by context.

}

In this paper, we propose a novel method named Narrative Player, which aims for the \vis{enhancing the reading experience by} automatic revival of data narratives with visuals. 
\vis{It generates animated visual sequences with audio and subtitles to provide users with a multi-channel experience. For instance, for the data narratives presented in Fig.~\ref{fig:teaser}, Narrative Player will generate the animated visuals with transitions and convey textual information through audio narration and subtitles, thereby enhancing the user experience.} 
Narrative Player consists of two major modules, i.e., narrative analysis for analyzing insights within the whole paragraph and visual generation for creating vivid visual stories for the paragraph.
Given a data narrative and its corresponding data table as input, the narrative analysis module utilizes large language models (LLMs) to extract contextualized data facts (i.e., numerical or statistical results of data mining~\cite{wang2019datashot}) from the narrative, representing the underlying story of the narrative.
The visual generation module then organizes the extracted contextualized data facts into a cohesive and contextually meaningful visualization sequence with a carefully designed optimization function.
Additionally, the module incorporates seamless transition animation effects between adjacent visualizations to strengthen user engagement.
Finally, the visualization sequence, transition animations, and audio narration generated by text-to-speech technologies are combined to produce a data video with narration-animation interplay.

To evaluate Narrative Player, we conducted a user study and expert study.
In the user study, we examined the data videos generated by Narrative Player \vis{against a set of baselines by user-perceived satisfaction.
\visrev{These baselines included} plain text to assess the enhancement of user experience, videos crafted by professionals to evaluate video quality, and videos from two ablation studies to measure the effectiveness of our two modules.
The results revealed that our generated data videos effectively enhance the user experience through two essential modules. Furthermore, these videos were well-received and displayed comparable quality to human-composed ones.
}
We also presented the videos generated by Narrative Player to experts for their feedback from a professional perspective, \vis{incorporating multidimensional ratings and interviews.}
The experts agreed that Narrative Player shows satisfactory performance in generating coherent visuals and maintaining consistency and contextualization, \vis{contributing to enhancing reading experience in both comprehension and enjoyment}. 

In summary, our contributions are concluded as follows:
\begin{compactitem}
    \item Narrative Player, an approach for automatically generating consistent, contextualized, and animated visual sequences for data narratives to enhance the reading experience.
    \item A narrative analysis module powered by LLM and embedding models for handling semantics and extracting data facts from long narratives, and a visual generation module powered by optimization considering \visrev{side-by-side} visualizations, visual focus, and primary visualizations to select a contextualized and consistent visual sequence. 
    \item An evaluation combining a user study for the analysis of user-perceived satisfaction, and an expert study \visrev{with ratings and interviews} to demonstrate the effectiveness of Narrative Player.
\end{compactitem}


\section{Related Work}
In this section, we discuss related works from three perspectives, i.e., natural language-based data visualization generation, visual generation, and text-driven storytelling.
\subsection{Natural Language-Based Data Visualization Generation}
Recently, natural language-based data visualization generation technologies have gained increasing attention, allowing users to express their intents conveniently using familiar natural language~\cite{Shen2021a}.
For example, commercial software such as Microsoft Power BI and Tableau both have their natural language interface (NLI) services.
Techniques range from translating natural language queries into visualization specifications~\cite{Narechania2020, Luo2021a, Cui2020b} to enhancing user interaction through natural language and multimodal inputs~\cite{Yu2020b, Srinivasan2020a, Srinivasan2018}. 

In addition to one-shot visualization generation, recent studies have explored conversational interaction experiences~\cite{Tory2019b, Shen2021a, Setlur2022}. 
Some studies enhanced the efficiency and effectiveness of NLIs. They have explored applying pragmatics principles to analytical dialogues~\cite{Hoque2018, Srinivasana}, simplifying query processes~\cite{Dhamdhere2017}, and providing explanations for understanding visualization outcomes~\cite{Feng2023}. Others expanded the functionality and interactivity of NLIs. They have developed interactive dialogue systems for direct visualization interaction~\cite{Setlur2016}, complex request and simultaneous conversations handling~\cite{Fast2017, Mitra2022}, and visualization authoring~\cite{vistalk}, alongside exploring design principles for NLI-based data exploration~\cite{Tory2019b,Setlur2022}. 

Overall, the aforementioned systems mostly generate visualizations based on simple natural language queries (e.g., ``show me how horsepower varies each year by origin'') for data analysis purposes. However, there is a lack of work that generates a sequence of visualizations for visual data storytelling purposes based on \vis{the context of} data narratives. Our work extends the scope of natural language-based data visualization generation technologies by moving towards augmenting data narratives with dynamic data charts for storytelling.

\subsection{Visualization Sequence Generation}
Narrative sequencing plays a vital role in visual data storytelling, which can affect the audience's comprehension.
Hullman et al.~\cite{Hullman2013a} conducted two studies to identify a relative ranking of visualization transition types by viewer perspective.
GraphScape~\cite{Kim2017b} further formulates a directed graph-based model for measuring visualization similarity and ranking visualization sequences.
Based on the model, TaskVis~\cite{shendata} proposes a task-driven strategy to recommend a set of combined visualizations to give an overview of the dataset.
\visrev{Dziban~\cite{lin2020dziban} supports both partial specification visualization and an anchoring mechanism for conveying the desired context.}
Shi et al.~\cite{Shi2019} leverage reinforcement learning to generate chart sequencing, capturing the relationships between charts.

Data facts, derived from raw data, underpin analysis and understanding, while visual data stories enlive these facts with sequential visualizations for easier interpretation.
For instance, DataShot~\cite{wang2019datashot} can automatically generate fact sheets from tabular data, consisting of three parts, i.e., fact extraction, fact composition, and visual synthesis.
Calliope~\cite{shi2020calliope} accepts a spreadsheet as input, progressively generates story pieces, and organizes the facts into a visual data story based on the fact importance. 
AutoClips~\cite{Shi2021a} can automatically generate a data video by configuring a sequence of data facts with a fact-driven animation clip library.
Erato~\cite{Sun2022} supports human-computer cooperation to create visual data stories, where the user needs to provide a set of keyframes and describe the story theme and structure.
ChartStory~\cite{Zhao2023} is designed to help users craft comic-style data stories from a set of user-created charts, suggesting the underlying sequence of data-driven narratives.

Coherent visualization sequences can enhance readers' understanding and memory of data stories. However, most work creates sequences driven by user-inputted data facts or charts, neglecting the importance of narrative text with context. This paper aims to \vis{generate vivid data stories, specifically required consistency and contextualization, by understanding data narratives' context and generating coherent visuals controlled by context,} rather than directly configuring visuals or clips in data-driven ways.

\subsection{Text-Driven Storytelling with Visuals}
Narrative text plays an essential role in data-centric storytelling~\cite{Segel2010}, often in conjunction with visualizations to promote information dissemination.
For example, a set of works focuses on the efficient writing of data documents enriched in visualizations to present data insights~\cite{Latif2021, Sultanum2021, Chen2022a, DataParticles2023, Grimley2018, sultanum2023datatales}.
Aiming at enlivening news articles, Contextifier~\cite{Hullman2013} and NewsViews~\cite{El-Dairi2019} automatically generate annotated visualizations based on the news text to aid storytelling. \chirev{Elastic document~\cite{elasticdocuments} is an interactive system for generating on-demand visuals based on readers' selection of narratives and tables.} 
Most recently, Charagraph~\cite{Masson2023} helps paper readers to dynamically create interactive charts from data-rich paragraphs to obtain a better sense of data in text. 
\chirev{These works primarily employ keyword-matching techniques to select data for visuals and are only aware of the local context as the input limits to individual sentences or a few sentences as snippets. Advanced NLP technologies are not utilized to deal with extensive contextualization
, while we try to employ LLMs to infer the semantics of vague clauses in long narratives and fully convey the data story.}

Moreover, by enhancing static materials with compelling dynamics, data video has also been gaining increasing popularity in recent years.
Cheng et al.~\cite{Borgo2022} investigated the roles and interplay of narrations and animations in data videos, highlighting their combined role in storytelling.
Following the study, WonderFlow~\cite{wonderflow} is an interactive authoring tool that allows users to easily create data videos with narration-animation interplay from static visualizations and text.
Data Player~\cite{data_player} further automates the process. It first leverages LLMs to create text-visual links and then recommends an animation sequence based on constraint programming. \chirev{These works focus on animate single given visual under the guidance of narration, instead of generating visual sequences from solely data narratives.}

Despite the abundance of research on text-driven storytelling techniques, most of these methods require users to invest a significant amount of time and effort into creating stories with visuals and undergoing trial-and-error processes~\cite{Grimley2018, Conlen2021, DataParticles2023, Chen2022a, Masson2023,wonderflow} or ask users to provide complete narrative texts and data visualizations~\cite{Latif2021, Sultanum2021,data_player, Chi2020}. 
Our work is in line with the research of text-driven storytelling and attempts to automatically generate \chirev{consistent and contextualized visual sequences} based solely on data narratives to enhance the vividness of data stories.

\section{Narrative Player}
In this section, we first give an overview of Narrative Player and discuss the two key modules, i.e., narrative analysis and visual generation.

\subsection{Overview}

\subsubsection{Design Requirements}
\label{sec: design requirements}
\chirev{
Previous studies have enhanced reading materials with the integration of individual visuals~\cite{Hullman2013, elasticdocuments,Masson2023}. 
However, these enhancements partially address the complexities of comprehension, as data narratives encompass coherent insights that necessitate diverse visuals for a complete understanding~\cite{latif2021deeper}.
\vis{Additionally, they have not exploited enjoyment}. 
The importance of \vis{animated} visual sequences \vis{for experience enhancement} lies in their ability to present a memorable and complete story~\cite{jaffar2012youtube,niknejad2015comprehension}. Such sequences link individual data insights and contexts within the narrative \vis{through transitions and audio narration}.
This approach, however, poses challenges \vis{associated with ensuring consistency and contextualization}, surpassing the complexities of generating single visuals with limited context.
Acknowledging this, we have derived a set of design requirements based on the previous work \vis{and our motivation} to inform the design of Narrative Player.}


\textbf{R1 Understanding the narrative and extract textual intents:}
To \vis{present the data stories}, \tvcgminor{Narrative Player should first} analyze the input narrative to understand its underlying \visrev{explicit and implicit textual intents~\cite{Tory2019b}} and extract the key data facts for visual representation. 
The extracted intent information acts as the abstraction layer between the data narrative and generated visuals, ensuring that the visuals effectively communicate the intended message.

\textbf{R2 Handling ambiguous expressions within context:}
Due to the ambiguous nature of natural language, data narratives may contain expressions from which data facts cannot be extracted explicitly~\cite{Shen2021a}, but they are important to the coherence of the contextual story. Ignoring such sentences can lead to confusion for the viewer\vis{ and inversely harm comprehension and enjoyment}. To handle this, \tvcgminor{Narrative Player} should analyze the context in which an ambiguous sentence appears and determine the most likely interpretation based on surrounding information.

\textbf{R3 Aligning textual and visual intents:}
A seamless binding between the narrative and visuals \vis{is crucial for user experience, necessitating a close alignment between} the textual and visual intents~\cite{data_player, DataParticles2023,Borgo2022}. This may involve using similar visual elements or themes in the visualizations to reinforce the key ideas or concepts presented in the text. Additionally, \tvcgminor{Narrative Player} should be able to adapt the visuals to match better the intended message, such as changing the color or shape of visualizations to emphasize a particular point in the narrative~\cite{Tory2019b}.

\textbf{R4 Ensuring visual consistency \visrev{and contextualization}:}
Consistency in style, color, and layout is crucial for cohesive visual representations that match the narrative's theme and tone.
In addition, visual sequences should balance visualization diversity and user cognitive load to help maintain the viewer's understanding of context. Frequent changes in the sequences can impose a huge cognitive load on the viewer~\cite{data_player}. Real-world data videos typically center on one or two main visualizations, supplemented by other embellished information~\cite{Borgo2022}.

\textbf{R5 Enabling \vis{multi-channel perception with transitions, audio and subtitles}:}
Static rendering of visual sequences can bore viewers and obscure differences between visuals. However, dynamic transitions between different visualizations can effectively engage viewers and clarify narrative progression~\cite{kim2020gemini}. \tvcgminor{Narrative Player} should enable smooth transitions between visualizations by incorporating transitional animation effects (e.g., fade-ins, wipes, or slides) that enhance the flow of the narrative.
\vis{Audio adds dimension to the display, with narration acting as subtitles to guarantee that the final video accurately conveys the original textual information.}

\subsubsection{Workflow}
\begin{figure}[t]
  \centering
  \includegraphics[width=\columnwidth]{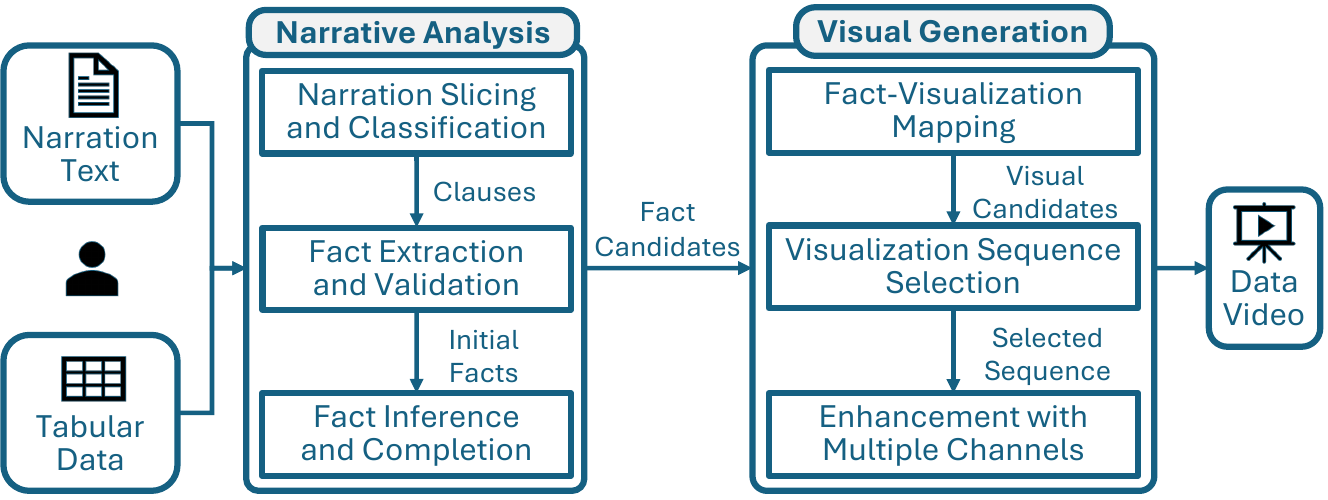}
  \caption{\tvcgminor{Narrative Player system overview and processing pipeline. The Narrative Analysis module automatically processes user-provided narration text and tabular data to generate \tvcgminorr{fact candidates which} the Visual Generation module then uses to create data videos.}}
  \label{fig:pipeline}
  \vspace{-8px}
\end{figure}

\visrev{Based on the aforementioned design requirements, we propose Narrative Player to automatically revive data narratives with visuals. It works sequentially, beginning with the extraction of contextualized data facts, which serve as elementary building blocks of a data story. This is followed by the generation of visuals, resulting in a well-organized visual sequence and multi-channel experience via videos. Fig.~\ref{fig:pipeline} illustrated the two key modules used in this pipeline:}

\visrev{
\textbf{Narrative Analysis}. The narrative analysis module (Sec.~\ref{sec: narration analysis}) divides the narrative text into clauses\footnote{\visrev{According to English grammar scholars~\cite{winter2020towards}, a clause is identified as a unit of a sentence that forms recognizable syntactic constituents: a subject and predicate, with or without adjuncts, or a predicate with or without adjuncts.}} by the LLM \tvcgminor{(Sec.~\ref{sec: narration slicing and classification})}, which is viewed as the basic units that contain data facts \tvcgminor{(Sec.~\ref{sec: fact formulation})}. Then it uses the LLM with well-designed prompts and the Sentence Embedding model to extract and validate data facts from clauses (R1) \tvcgminor{(Sec.~\ref{sec: extraction and validation})}. Initially, this yields validated facts that are semantically aligned with the narrative, which are a subset of all fact candidates. The module then employs LLM and rule-based heuristics to infer and complete fact sets (R2) \tvcgminor{(Sec.~\ref{sec: disambiguation within context})}, ultimately producing enriched, contextualized data facts.
}

\visrev{
\textbf{Visual Generation.} The visual generation module (Sec.~\ref{sec: visual generation}) maps the contextualized facts candidates into visualization candidates (R3) \tvcgminor{(Sec.~\ref{sec: visual mapping})}. Leveraging an optimization function emphasizing both contextualization and consistency (R4), it determines a specific visualization for each clause, crafting a coherent sequence \tvcgminor{(Sec.~\ref{sec: visual sequence selection})}. 
Furthermore, to enable a more engaging storytelling experience (R5), the module enhances the visualization sequence with transition animations and audio narration and finally renders them into a data video \tvcgminor{(Sec.~\ref{sec: enhance with multiple channels})}.
}

\subsection{Narrative Analysis}
\label{sec: narration analysis}
To generate corresponding visual aids from data narratives, we need to 
\vis{understand the narrative intents by extracting data facts (R1) and handle ambiguity within the context (R2).}
\vis{The task is complex with technical challenges (TCs). Specifically, data facts must be contextualized to align with the narrative's semantics (TC1) and its broader context (TC2) and must ensure completeness to enrich visual generation (TC3).}

\visrev{Several conversational NLI systems consider the context of the preceding NL utterance by keywords-parsing~\cite{Tory2019b} or LSTM~\cite{vistalk}, but they can't address the wider context of the whole narration.}
\visrev{Prior text-driven storytelling works also use keywords-parsing~\cite{elasticdocuments, Masson2023,sultanum2023datatales}} and fall short in structuring data fact representations from lengthy narratives.
They mostly operate on individual words or phrases without adequately addressing their relations \visrev{or handling ambiguities}. 
\vis{Besides, the lack of sufficient datasets for training in data storytelling has long been acknowledged~\cite{lu2021automatic}.
LLMs excel in context-rich tasks and scenarios with sparse resources~\cite{wang2023gptner} due to their ability in few-shot learning and human-like reasoning~\cite{li2023survey}.
Hence we design the automatic narrative analysis process based on LLMs.}
\visrev{We use the OpenAI GPT-4-turbo model with 128k context and temperature 0.3.}

We will detail the preparation and definitions in Sec.~\ref{sec: narration slicing and classification} and Sec.~\ref{sec: fact formulation}, describe the methods for fact extraction and validation in Sec.~\ref{sec: extraction and validation} (TC1), and address issues of context-based inference (TC2) and fact completion (TC3) in Sec.~\ref{sec: disambiguation within context}. 

\subsubsection{Narration Slicing and Classification}
\label{sec: narration slicing and classification}
Data narratives consist of two main components: factual sentences that present data facts and story or background sentences that provide context~\cite{Borgo2022}.
These two segments may be interspersed with visualizations to enhance understanding.
To understand the structure of the data story, narratives are first segmented into sentences using punctuation marks and LLMs assess each sentence for data facts presence. Factual sentences are further segmented into clauses for subsequent processing.

\subsubsection{Data Fact Formulation}
\label{sec: fact formulation}
One sentence can have multiple clauses that correspond to various data facts~\cite{Borgo2022}. We consider clauses as the basic units of the narrative for data fact extraction. 
We formalize the data fact as a 6-tuple structure\visrev{, adapted from DataShot~\cite{wang2019datashot}, but exclude the importance score as we derive facts matched to narrative semantics rather than mining them from tabular input, while we discuss the future consideration of fact importance at the end of Sec.~\ref{sec: discussion}}.

\vspace{-16pt}
\begin{equation}
\scalebox{0.8}{$fact:=\{type, parameters, measure(s), context, breakdown(s), focus\}$}
\end{equation}
\vspace{-16pt}

where \textit{type} denotes the data fact type (e.g., trend, comparison, deviation), \visrev{as adopted from TaskVis~\cite{Shen2021}}, with \textit{parameter} specifying its details like deviation value; \textit{context} defines the data subspace, \textit{breakdowns} segment this subspace into groups, and \textit{measure} assess each group's value, with \textit{focus} highlighting specific data\visrev{, as shown in Fig.~\ref{fig:rewriting}}. 
On this basis, we define the whole narrative structure that reflects the underlying story as a dictionary, i.e.,
$story: \{ clauses[i] : facts_i \mid i = 1, 2, \ldots, n \}$, where \(clauses[i]\) represents the \(i\)-th clause in the data narrative, \(facts_i\) is the set of all facts related to \(clauses[i]\), and \(n\) is the total number of clauses. Each \(facts_i\) can be defined as: $facts_i = \{facts[i,1], facts[i,2], \ldots, facts[i,m_i]\}$, where \(facts[i,j]\) signifies the \(j\)-th \textit{fact} conveyed in the \(i\)-th clause and \(m_i\) is the number of facts for the \(i\)-th clause.

\subsubsection{Data Fact Extraction and Validation}
\label{sec: extraction and validation}
\begin{figure}[t]
  \centering
  \includegraphics[width=\columnwidth]{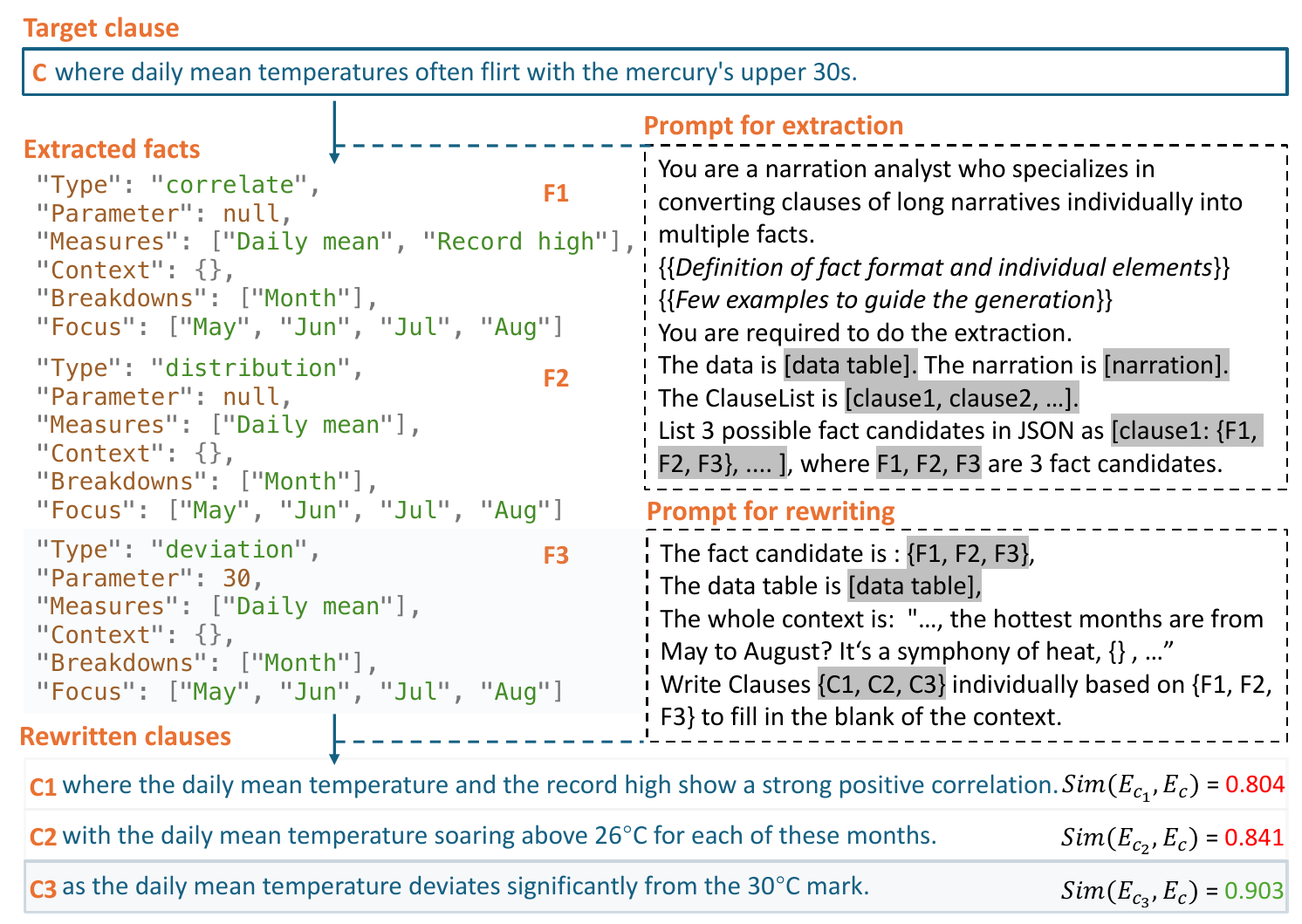}
  \caption{\tvcgminor{An illustrating case describing how to extract, validate, and select data fact candidates. In one LLM session for preliminary extraction, three facts (F1, F2, F3) have been extracted for one clause (C). Another LLM session with appropriate prompts rewrites the target clause based on the three facts and context into three rewritten clauses (C1, C2, C3). The sentence embedding model generates embeddings ($E_{c_{i}}$ and $E_{c}$) for all the clauses. Finally, F3 was ranked first based on the cosine similarity.} }
  \label{fig:rewriting}
\end{figure}
By providing the model with well-composed prompts and examples \chirev{as shown in the lower left of Fig.~\ref{fig:rewriting}}, we effectively guide models in extracting data facts behind each clause 
\vis{(TC1)}. 
When \visrev{prompting} LLM, considering its hallucination nature~\cite{ji2023survey}, we first try to achieve a comprehensive data fact set for each clause by utilizing three sessions, each using the same prompt \visrev{with temperature 0.3} and generating three fact candidates per clause. This approach yields nine potential data facts for each clause.

From analyzing clause-data fact relationships, we identify that generated candidates may overlap or diverge from the text's semantics (F1, F2, F3 in Fig.~\ref{fig:rewriting}), necessitating validation and filtration to ensure the data facts' accuracy and reliability.

We proposed a novel method that leverages LLMs and sentence embedding for data fact candidate validation. 
Specifically, as shown in Fig.~\ref{fig:rewriting}, for a clause (C) and a specific data fact candidate (F3), we dig the clause out of the narrative \chirev{as context} and fill in new clause (C3) based on \chirev{this context} and the data fact (F3).
The rewritten clause tends to be more direct in its representation of data facts, omitting the vague expressions and rhetorical language often found in vague clauses. 
Then we use rankcse~\cite{liu-etal-2023-rankcse}, \visrev{the SOTA model for unsupervised sentence representation learning via learning to rank}, to obtain the embeddings of both the original and rewritten clauses. 
Finally, we calculate the cosine similarity between these two embeddings. As shown in Fig.~\ref{fig:rewriting}, C3 and C exhibit a high degree of similarity in their sentence embeddings, and thus F3 is re-ranked as the most qualified fact among these three. \chirev{This approach avoids using LLMs for self-correction, acknowledging their limitations in semantic understanding and sentence embedding~\cite{muennighoff2022sgpt}.}

\chirev{In our observation}, clauses with clear data properties or easily inferred values often closely match several rewritten clauses. Thus ``clear clause'' is defined as having at least 6 out of 9 candidates with a similarity score above 0.85, a threshold determined from empirical testing across narratives.
We remove duplicates and select the top three candidates by similarity as qualified data facts. 
Conversely, a ``vague clause'' produces fewer than 6 facts with scores above 0.85.

\subsubsection{Data Fact Inference and Completion within Context}
\label{sec: disambiguation within context}

After \visrev{prioritizing explicit intent~\cite{Tory2019b}} by data fact validation and clause characterization, two issues still exist for us to address \visrev{on implicit intents}: 
1) Vague clauses ambiguously reference data properties, values, or table subspaces. For example, ``the chill of winter finds its way'' might refer to various temperature metrics and specific months in the data table, without clarity.
2) Clear clauses yield three facts, yet these represent just a subset of all possible facts. For instance, ``especially in May and June'' might refer solely to these months or highlight their distinctiveness, suggesting varied interpretations as either a complete \textit{context} or specific \textit{focus} within a broader narrative.

\begin{figure}[t]
  \centering
  \includegraphics[width=\columnwidth]{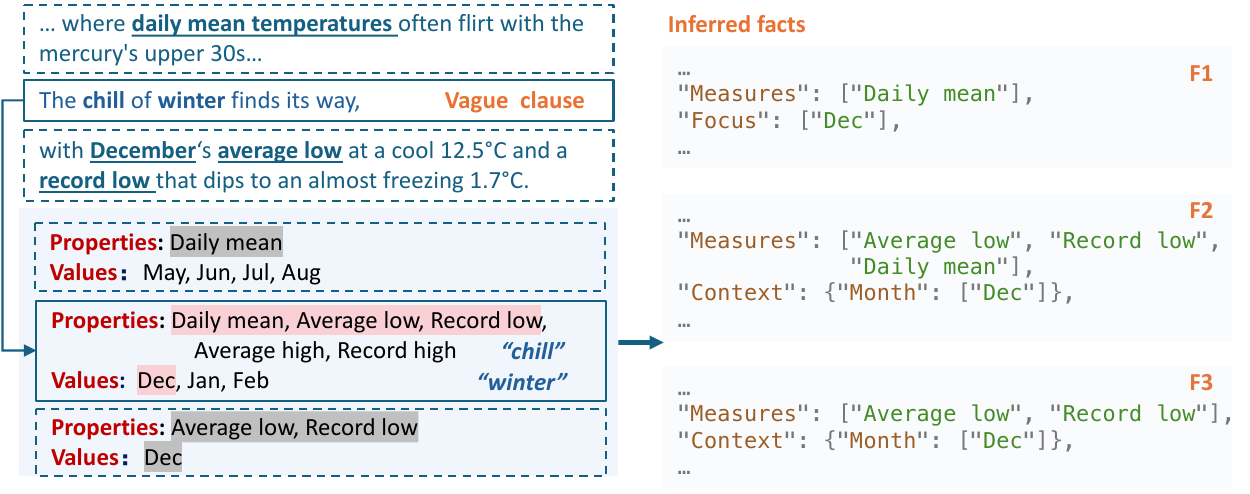}
  \caption{\visrev{An illustrating case \visrev{of inferring} vague clause within context. The narrative analysis module first detects keywords, ``chill'' and ``winter'', relating them to five candidate properties about temperature and three candidate values for months \visrev{based on the narration and data}. Then the module selects two clear clauses around as references, inferring two sets of referred properties and values, as shown in dashed boxes and arrows. The module further merges the intersection of them as filtered properties and values and extracts three fact candidates (F1, F2, F3). }
  }
  \label{fig:inference within context}
\end{figure}

The challenge with vague clauses is determining relevant data properties or values that match the clause semantics and context \vis{(TC2)}. To tackle this, we use another LLM session as outlined in Fig.~\ref{fig:inference within context}:

    Firstly, keywords are identified within each vague clause and mapped to potential data properties or values in the table. This process generates a candidate set of \textit{measure}, \textit{context}, or \textit{focus} for fact candidates. For instance, ``chill'' and ``winter'' lead to inferring five properties about temperature and three \visrev{context-appropriate} values about month \visrev{in the Northern Hemisphere, based on the location mentioned in the narration.}
    Then we infer reference data properties or values within context. Clear clauses that are semantically related and adjacent are selected as reference clauses. The intersection of the candidate set and the qualified candidate facts for reference clauses forms a set of reference data properties or values. In Fig.~\ref{fig:inference within context}, two adjacent clear clauses serve as references. The intersection of data properties or values—one property from the first and two properties plus one value from the second—are selected to match with the target vague clause's inferred properties or values.
    After that, the union of the intersection of each reference set and the candidate set establishes the filtered data properties or values. In Fig.~\ref{fig:inference within context}, one value and three properties from the two references are merged as the results.  
    Finally, the fact candidates are generated. If no candidates are initially identified, fact candidates from the nearest clause are used. Otherwise, three fact candidates are generated using a subset of the filtered properties or values as \textit{measure}, \textit{context}, or \textit{focus}. Each candidate aligns semantically with the first reference clause (F1), the second reference clause (F3), or a combination of both (F2), as shown in Fig.~\ref{fig:inference within context}.

The process of obtaining a complete fact set \vis{(TC3)} involves addressing three types of incompleteness: single facts, multiple facts within a single clause, and multiple facts across clauses.
These are tackled by applying the following heuristics, \visrev{and use an Example Clause (EC: ``with December's average low at a cool 12.5°C and a record low that dips to an almost freezing 1.7°C.'') to illustrate the process.}.


For a single fact, since keywords represent either an entire subspace as \textit{context} or a \textit{focus} within a larger space, the fact candidates are expanded by interchanging these two elements, provided that the \textit{focus} corresponds to a smaller data range than the \textit{context}. 
\visrev{For example, the initially generated fact (noted as F) of EC includes the attributes: \{\textit{type}: \textit{extreme}, \textit{measures}: \{\textit{Average low}, \textit{Record low}\}, \textit{context}: \{\}, \textit{focus}: \textit{Dec}\}.
By swapping \textit{context} and \textit{focus}, $\text{F}_{1}$ is obtained, which includes the attributes: \{\textit{focus}: \{\}, \textit{context}:\{\textit{Month}:[\textit{Dec}]\}\}.
}

Also, fact candidates with the same \textit{measure} and \textit{breakdowns} but different \textit{type} and \textit{parameter} may be generated, suggesting similar semantics but arbitrary distinctions. In such cases, \textit{type} and \textit{parameter} are cross-combined between the two facts.
\visrev{Among the initial facts of EC, one has the same \textit{measures} and \textit{breakdown} as $\text{F}_{1}$, but its \textit{type} is \textit{distribution}. Replacing the \textit{type} of $\text{F}_{1}$ with \textit{distribution} results in $\text{F}_{2}$.}

Besides, keywords in clear clauses may capture part of the context but miss the context suggested by adjacent clauses. If a fact matches an adjacent clause's fact in either \textit{measure} or \textit{context} and is a subset of that fact, we expand the \textit{measure} or \textit{context} to match.
\visrev{
The \textit{measures} of a fact in the previous sentence of EC, ``The chill of winter finds its way'', also include \textit{Daily mean} (see Fig.~\ref{fig:inference within context}). Adding it to the \textit{measures} of $\text{F}_{2}$ results in $\text{F}_{3}$.
$\text{F}_{3}$ has the attributes: \{\textit{type}: \textit{distribution}, \textit{measures}: \{\textit{Average low}, \textit{Record low}, \textit{Daily mean}\}, \textit{context}:\{\textit{Month}:[\textit{Dec}]\}, \textit{focus}: \{\}\}.
$\text{F}_{3}$ differs significantly from F but aligns closely with the context.
}

We observe that the rewritten text of fact candidates after completion closely matches the original. The \visrev{generated and refined facts indicate the intent comprehension and} supports the further visual generation.


\subsection{Visual Generation}
\label{sec: visual generation}

Contextualized data facts, bridging narratives and visuals, outline the story structure within data narratives. 
\vis{To deliver a vivid experience, Narrative Player then maps data facts to visuals, organizes them into sequences, and synthesizes videos that incorporate animated transitions, audio, and subtitles across multiple channels. }

\subsubsection{Data Fact-Visualization Mapping}
\label{sec: visual mapping}
\begin{figure*}[ht]
  \centering
  \includegraphics[width=\textwidth]{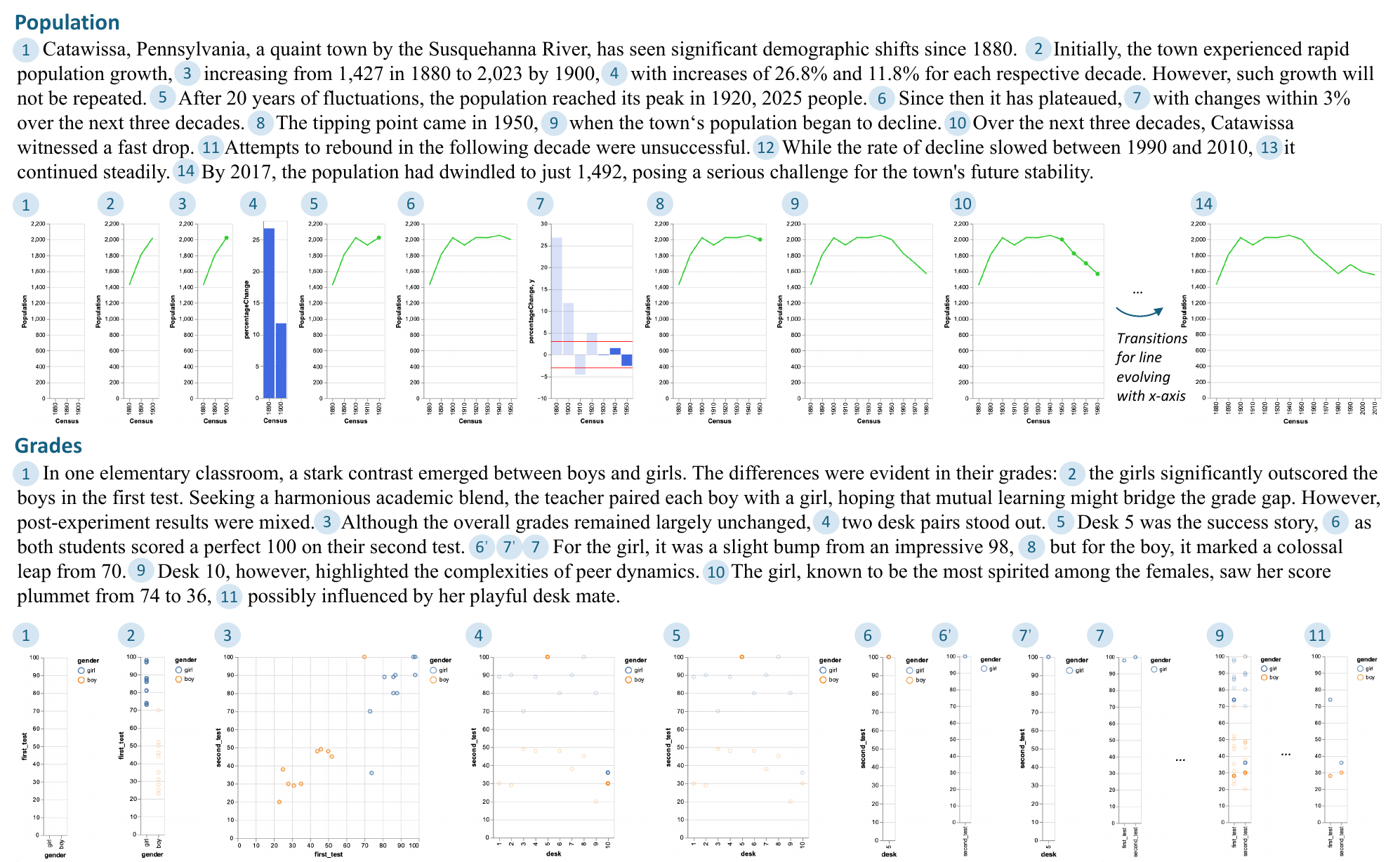}
  \caption{\tvcgminor{Two examples automatically generated by Narrative Player, each featuring a visualization sequence with corresponding transition animations. Transition animations are activated when the audio narration hits the relevant segments, with representative transitions highlighted in the text. More examples can be found \tvcgminorr{on} \url{https://datavideos.github.io/Narrative_Player/}.}}
  \label{fig:examples}
  \vspace{-10pt}
\end{figure*}

\vis{Inspired by existing mapping relations between data, task, and visual marks~\cite{shendata, wang2019datashot}, we} map each data fact candidate to the Vega-Lite visualizations~\cite{satyanarayan2016vegalite} based on their \textit{type}, \textit{breakdowns}, and \textit{measure}. Narrative Player emphasizes \textit{focus} of data facts by adjusting the opacity and stroke of relevant visual elements.
\chirev{For example, the data table of ``Grades'' in Fig.~\ref{fig:examples} includes four attributes: $\{gender, first\_test, second\_test, desk\}$. The partial fact of 2 is $\{type: comparison, measures: \{first\_test\}, breakdowns: \{gender\}\}$, while fact of 3 differs by having $\{measures: \{first\_test, second\_test\}\}$. 
This distinction influences the x and y-axis settings in visuals 2 and 3, where the selection of visualization data strictly follows the fact, assuming pre-processed input tables. Currently, we support basic types such as (stacked, grouped) bars, (multi-series) lines, points, and ticks.
}
In line with real-world data video practices \visrev{and the prior empirical study~\cite{Tory2019b}}, we incorporate \visrev{side-by-side visualizations} for the ``compare'' and ``correlate'' fact types, using \textit{breakdowns} as the aligned axis, displayed either vertically or horizontally.

Narrative Player enhances visualization sets for consistency between text and visuals, drawing on Qu and Hullman's~\cite{qu2017keeping} consistency principles for Multiple Views visualization. 
Narrative Player standardizes the color and scale across charts in these sets. 
It standardizes color and scale across charts by extracting fields from visuals and using LLM to match comparable fields. 
LLM recommends color palettes \visrev{that align with the semantics~\cite{Tory2019b}}, and uniformity in color, scale, and order is ensured based on whether fields are the same or not.
As illustrated by visual 10 and visual 11 in \chirev{Fig.~\ref{fig:teaser}, they share the same y-axis range and use varying shades of red to differentiate the three comparable temperature fields semantically}. 
Moreover, it applies consistent visual effects, like stroke or opacity, to highlight story patterns (Fig.~\ref{fig:examples}).

Additionally, for vague clauses, Narrative Player adjusts visualizations to better align with the narrative \chirev{(R3)}. 
For vague clauses initiating a sequence, visual emphasis is removed to prevent premature data insights, as shown in visual 2 of \chirev{Fig.~\ref{fig:teaser}}; otherwise, it mimics the previous visualization. For clauses lacking data facts, mid or end clauses maintain the prior visualization, while starting sentences adopt features from the next one, keeping only the axes and title.

\subsubsection{Visualization Sequence Selection}
\label{sec: visual sequence selection}
After mapping data facts to visuals for each clause, we select one visualization per clause from the candidates to create a sequence that represents the data story.
Narrative Player selects the optimal visualization sequence $V = (V_{0}, V_{1}, V_{2}, ..., V_{n})$ from a large set, where $V_{0}$ is a $null$ spec for specific-to-general selection~\cite{Kim2017b}, and other $V_{i}$ are clause-specific visualizations.
\visrev{Similar to prior works~\cite{Shi2021a,Zhao2023}, we adopt a global optimization of the visual sequence, instead of incremental methods adopted in the conversational NLI system~\cite{Tory2019b}, to ensure}  dynamic consistency and contextualization \chirev{(R4)}. Inspired by literature~\cite{Hullman2013a, Kim2017b, Shi2021a, ACTR}, Narrative Player considers three heuristic features: 1) the cost of dynamic transitions \chirev{of \visrev{side-by-side} visualizations and fixed elements} for maintaining local similarity and consistency, 2) the emphasis on visual focus for insight understanding, and 3) the activation of the primary visualization for global contextualization.

    \textbf{Minimize the dynamic transition cost \chirev{by considering \visrev{side-by-side} visualizations and fixed elements}.} \visrev{We use the transition cost model from Graphscape~\cite{Kim2017b} as a basic operator, which has been widely used in prior studies~\cite{Shen2021,shi2020calliope,Zhao2023,lin2020dziban} for measuring visualization similarity and ranking visualization sequences.} Narrative Player further considers 1) \visrev{side-by-side} visualizations and 2) fixed elements in dynamic visuals to ensure local visual similarity. 
    By minimizing the sum of transition costs $\mathcal{T}$ as follows, adjacent clauses that align with similar visualizations within their semantic scope are ensured, thereby addressing the challenge of verifying facts for vague clauses.
    
    \vspace{-10pt}
    \begin{equation}
    \mathcal{T} = \sum\limits_{i=1}^{|V|} T''\left(V_{i-1}, V_{i}\right)
    \end{equation}
    \vspace{-10pt}
    
    where $T''\left(V_{i-1}, V_{i}\right)$ is the transition cost variants combination between two adjacent visualizations. 
    \visrev{Given that a} clause may link to \visrev{side-by-side visualizations} when the fact \textit{type} is either ``compare'' or ``correlate'',
    we denote $V_{i-1}=\{s_1,s_2\}$ and $V_{i}=\{e_1,e_2\}$, where $s_{i}$ and $e_{j}$ are individual visualizations extracted from $V_{i-1}$ and $V_{i}$, respectively. This results in four possible transition scenarios: one-to-one, one-to-two, two-to-one, and two-to-two:
    
    \vspace{-10pt}
    \begin{equation}
    \label{eq.4}
    \scalebox{0.68}{$
    T''(s, e) = 
    \begin{cases} 
    T'(s_{1}, e_{1}) & s_{2} = null, e_{2} = null \\
    T'(s_{1}, e_{1}) + T'(s_{1}, e_{2}) & s_{2} = null, e_{2} \neq null\\
    T'(s_{1}, e_{1}) + T'(s_{2}, e_{1}) & s_{2} \neq null, e_{2} = null \\
    \min\left( T'(s_{1}, e_{1}) + T'(s_{2}, e_{2}), T'(s_{1}, e_{2}) + T'(s_{1}, e_{1})\right) & s_{2}\neq null, e_{2} \neq null
    \end{cases}$}
    \end{equation}
    \vspace{-10pt}
    
    where $T'(s_{i}, e_{j})$ is the variant of static visualization transition cost for one-to-one transition. For a one-to-two or two-to-one transition, we sum the costs of two pairs of visualizations. For a two-to-two transition, the minimum sum of the intersecting transition costs is chosen, as each visualization from the initial clause moves independently to the next, simplifying the need for comprehensive combinations \chirev{(see visuals 8, 9, and 10 in Fig.~\ref{fig:teaser})}.
    The one-to-one transition computation differs from the standard static visualization transition cost $T(s_{i}, e_{j})$:
    \vspace{-8pt}
    \begin{equation}
    \scalebox{0.72}{$
    \label{eq.5}
    T'(s_{i}, e_{j}) = 
    \begin{cases} 
    T(s_{i}, s'_{i}) + T(e_{j}', e_{j}) & \text{if } s_{i} \text{ and } e_{j} \text{ have varied fields with join keys} \\
    T(s_{i}, e_{j}) & \text{otherwise}
    \end{cases}$}
    \end{equation}
    \vspace{-8pt}
    
    The interim states $s_{i}'$ and $e_{j}'$ of $s_{i}$ and $e_{j}$, respectively, limit the data range to their shared data, as all other visual aspects stay the same. This method arises from noting that in dynamic visuals, maintaining specific elements like bars in joined visualizations minimizes inconsistencies due to data alterations
    \chirev{(see visuals 10, 10', 11', and 11 in Fig.~\ref{fig:teaser})}.

    \textbf{Emphasize the visual focus.} 
    Heuristic analysis suggests that a clear clause typically seeks to draw users' attention to key insights, often achieved through specific visual cues like opacity, stroke, or annotation. Hence, we design a visual focus bonus. 
    $\mathcal{B}$: 
    
    \vspace{-8pt}
    \begin{equation}
    \mathcal{B} = \sum\limits_{i=1}^{|V|} B(V_{i})
    \end{equation}
    \vspace{-8pt}
    
    If $V_{i}$ corresponds to clear clause and its \textit{focus} $\neq$ null, we set $B(V_{i})$ as 1; otherwise, it's 0. This ensures Narrative Player doesn't always favor visual sequences with a narrow data range lacking visual focus. 
    Whereas, vague clauses don't receive this consideration since they typically function as transitions between insights rather than highlighting specific visual focuses.
    
    \textbf{Activate the primary visualization.} Data videos often feature a primary visualization that either stays visible for long stretches or frequently returns, anchoring the topic and maintaining context. 
    \visrev{In computational psychology, models are proposed to measure the activation and retrieval probabilities of items in working memory. These models consider each item's exposure duration along with mechanisms of forgetting and interference, capturing how cognitive load and time influence memory dynamics~\cite{ACTR,TBRS,TBRS*}. Drawing from these models, }
    Narrative Player encodes the activation level of individual visualizations as $A=\{A_{1}, A_{2},..., A_{n}\}$, and encodes the retrieving probability of primary visualization as $\mathcal{P}$:
    
    \vspace{-18pt}
    \begin{gather}
    \mathcal{P}=\max\limits_{i}\frac{e^{A_{i}}}{\sum\limits_{j} e^{A_{j}}}
    \\
    A_{i}=\sum\limits_{k} (\alpha+\beta n_{i,k})
    \end{gather}
    \vspace{-12pt}
    
    where $n_{i,k}$ refers to the number of continuous clauses when the $i^{\text{th}}$ visualization appear for the $k^{\text{th}}$ time, while $\alpha$ and $\beta$ are two parameters for linear activation. 
    Due to the complexity of a fully realized computational model for working memory~\cite{TBRS,TBRS*}, we use a linear model, bypassing factors like forgetting and interference, to determine the primary visualization's prominence. Setting $\alpha$ at 1 and $\beta$ at 0.5 has proven effective in our context.

We aim to identify a visualization sequence that optimally satisfies all discussed factors. Thus, the overall optimization function $\mathcal{F}$ is defined as the weighted sum of three factors.

\vspace{-5pt}
\begin{equation}
\label{eq.10}
 \mathcal{F} = \omega_{1}\cdot\mathcal{T} +
 \omega_{2}\cdot\mathcal{B} +
 \omega_{3}\cdot\mathcal{P}    
\end{equation}
\vspace{-10pt}

Directly solving the optimization problem can result in extended computation times, especially with a large number of clauses generating numerous potential facts (e.g., exceeding 15 with each clause yielding between 3 to 8 potential facts). To mitigate this, we employed pruning strategies focused on maintaining visual consistency and insightfulness (R4), specifically when: 1) the same data field is represented in multiple visualization types, and 2) the sequence lacks a distinct visual focus. These strategies significantly cut down the generation time of visualization sequences to minutes, efficiently preserving performance.

\subsubsection{Enhance Visual Sequence with \vis{Multiple Channels}}
\label{sec: enhance with multiple channels}
\vis{Narrative Player provides a multi-channel enhanced experience by adding transitions to illustrate the visual story and maintain textual information through audio narration and subtitles.}

To achieve seamless transitions between narrative segments, we further apply transitional animation effects on the visualization sequence to guide the viewer through the sequence \chirev{(R5)}.
\visrev{We employ Gemini~\cite{kim2020gemini}, a system that combines declarative grammar with recommendations for animated transitions, to create the transition effects between individual visualizations.}
The overall transitions are determined by the transition cost calculation logic discussed in Eq.~\ref{eq.4} and Eq.~\ref{eq.5}:

\begin{compactitem}
    \item \textbf{No transition.} For adjacent visualizations $V_{i-1}$ and $V_{i}$ that don't share data fields and lack joined data, apply no transition effects.
    \item \textbf{One-to-One Transition.} 
    \visrev{We enable smooth transitions between $s_{i}$ and $e_{j}$ recommended by Gemini. Especially, when they} involve varied fields with joined keys as described in Eq.~\ref{eq.4}, we interpolate $s_{i}'$ and $e_{j}'$ with joined data between $s_{i}$ and $e_{j}$, as the transition between visual 10 and 11 in Fig.~\ref{fig:teaser}.
    \item \textbf{One-to-Two and Two-to-One Transitions.} The visualization pair with a smaller transition cost is the primary one to transition in Eq.~\ref{eq.5}. In a one-to-two scenario, after this primary transition, the secondary visualization emerges in the scene, as shown in Fig.~\ref{fig:teaser} when visual 8 transitions to 9. Conversely, in a two-to-one scenario, one visualization vanishes first, followed by the primary transition of the visualization pair.
    \item \textbf{Two-to-Two Transition.} The two selected visualization pairs, with the minimal transition cost combination as illustrated in Eq.~\ref{eq.5}, transition simultaneously.
\end{compactitem}

Furthermore, narration segments are converted into audio narration using text-to-speech technologies~\cite{text2speech}. By aligning the start and end times of each clause in the audio narration, Narrative Player ultimately renders the visualization sequence, transition animations, and audio narration into a data video with narration-animation interplay. \vis{Subtitles are also added and serve as another channel to present original text information for the reading experience.}

\section{Evaluation}

We evaluated Narrative Player's usability through three approaches. We conducted a user study \vis{illustrated the experience enhancement, video quality, and technical effectiveness of Narrative Player by comparing automatic-generated videos with plain text and those from ablation studies and human-made ones. We also gathered expert feedback through multidimensional ratings and interviews which provided in-depth insights and implications. Besides, real-world datasets were used to produce Data Videos, forming a gallery for illustration.}

\subsection{User Study}
This study \vis{1) analyzes the experience boost by videos against plain text, 2) assesses the quality of videos automatically generated against human-composed videos, and 3) evaluates the effectiveness of two main modules by ablation studies.}

\subsubsection{Dataset}

Our study uses datasets based on real-world narratives, including weather in Zhaoqing (``Weather''), the population in Catavissa (``Population''), COVID-19 regional patterns (``COVID-19''), GDP of top economies (``GDP''), course grades (``Grades''), and corporate sales (``Sales''). The ``Weather'' and ``Population'' data come from the ToTTo dataset~\cite{parikh2020totto} tailored for table-to-text tasks and are enhanced by an expert for richer storytelling. ``COVID-19'' data is from the WTO and also features in the CrossData~\cite{Chen2022a}, ``GDP'' from the World Bank and its affiliated blogs, and ``Grades'' and ``Sales'' are inspired by online data stories~\cite{grades-source, sales-source}, all refined to align with narrative styles.

\subsubsection{Preparation} \label{sec: preparation}
To evaluate the effectiveness of Narrative Player and its two main modules, we prepared five storytelling versions for each narrative: an automatically generated data video by Narrative Player (e.g., Fig.~\ref{fig:teaser} and Fig.~\ref{fig:examples}), plain text \vis{for studying experience boost \tvcgminor{following existing research~\cite{ying2023reviving} with similar study setting}}, a human-crafted data video by visualization researchers \vis{for assessing quality}, and two data videos from ablation studies \vis{for learning technical nuances}.
\chirev{Since it's a new and challenging task, there is no previous baseline available for comparison, and other rule- or template-based alternatives cannot be quickly designed to satisfy the design requirements in Sec.~\ref{sec: design requirements}.}

\textbf{Manual-composed by human.}
We enlisted two experienced visualization researchers, each with over three years in the field and publications in significant conferences like IEEE VIS and ACM CHI, to craft visual sequences and videos. Their selection was based on their proficiency with Vega-Lite for visualization and their background in animation and data video creation, ensuring that the designs were practical and grounded, thus avoiding comparisons with unfeasible, overly imaginative concepts. To guarantee a fair comparison, they were initially briefed on Narrative Player's core functionalities and design space. They then outlined their visual designs for specific narrative segments, either verbally or through sketches, which our programmer implemented. Through iterative refinement with these researchers, we ensured the final visual sequences and videos aligned with their visions.

\textbf{Ablation study on narration analysis.} 
This ablation study, termed ``Ablation1,'' assessed the narrative analysis module's effectiveness in extracting contextualized data facts. We applied the module to focus solely on clear clauses, as described in Sec.~\ref{sec: extraction and validation}, using these facts as anchors. Facts from adjacent vague clauses were treated as replications of these clear anchor facts, leading to visuals generated without the detailed inference and completion of vague facts within context.

\textbf{Ablation study on visual generation.} 
In this ablation study, termed ``Ablation2,'' we aim to validate the advantages of including visual focus and primary visualization in the optimization function for generating visualization sequences. We focused solely on the first term of Eq.~\ref{eq.10}, which involves minimizing the transition cost and has been empirically proven to be reasonable and necessary for creating data videos~\cite{Shi2021a}.

\begin{figure}[t]
  \centering
  \includegraphics[width=0.85\linewidth]{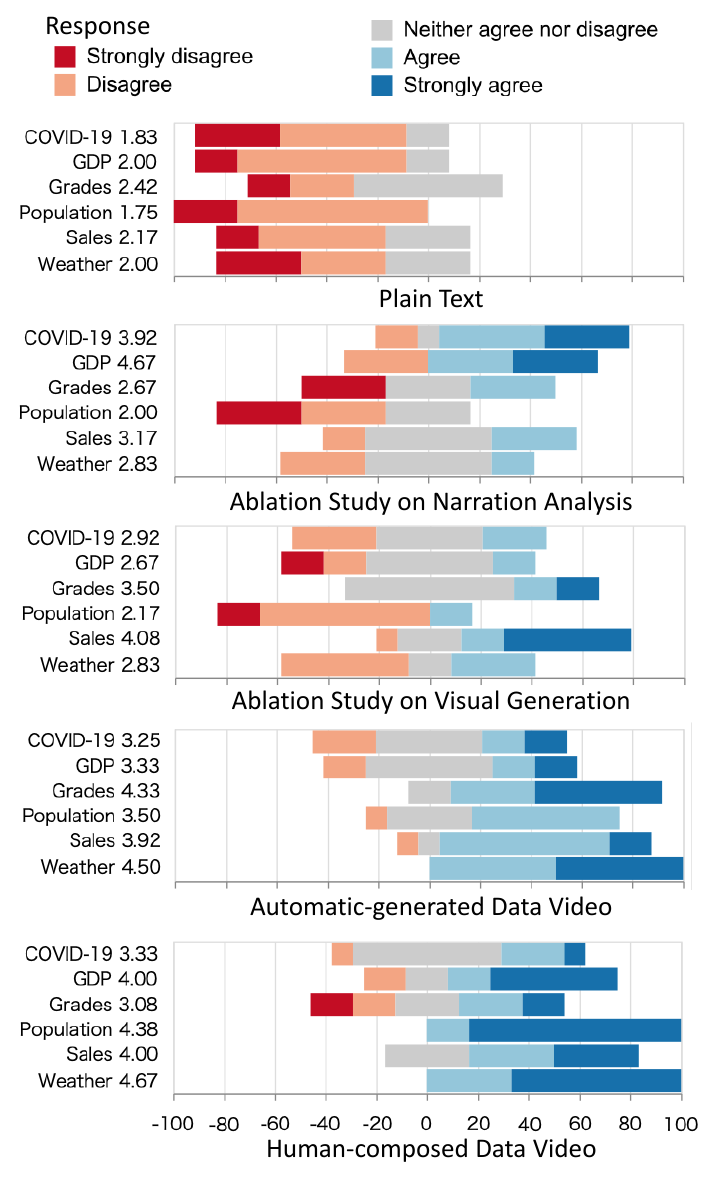}
  \caption{\tvcgminor{User ratings for general satisfaction with the five materials on a 5-point Likert scale, with means shown beside the bars.} } 
  \label{fig:ratings}
  \vspace{-10pt}
\end{figure}

\subsubsection{\tvcgminor{Participants and Procedure}}
We enlisted 12 participants (7 males, 5 females) \tvcgminor{aged between 20 and 29 years old ($M$=23.17, $SD$=2.58)}, labeled P1-P12, with backgrounds in reading data-rich documents, including data analysts and graduate students in fields like visualization (VIS), machine learning, data science, software engineering, human-computer interaction (HCI), and design. 
The study \visrev{was conducted with some participants attending in-person and others virtually.}
Each participant received six sets of materials comprising \visrev{data tables and five versions of storytelling versions (data narratives and four videos per narrative), as listed in Sec.~\ref{sec: preparation}}. 
The order of narratives and videos was randomized. 
Participants reviewed the narratives and tables before watching the videos \visrev{on their own computers}.
\tvcgminor{They rated each version on a 5-point Likert scale for general satisfaction \visrev{via Google Forms}. 
We leave more detailed ratings and analysis to the expert study (Sec.~\ref{sec: expert study}) and \tvcgminorr{discuss} more reliable experimental settings in Sec.~\ref{sec: discussion}.}
They could re-watch videos and were encouraged to share their thoughts openly throughout the 60-70 minute study. 
Participants received a \$15 for their time.

\subsubsection{\chirev{Qualitative Analysis of Perceived Satisfaction}}
\tvcgminor{We report our qualitative analysis results and findings below.}

\vis{\textbf{
Videos engage readers more effectively than text, with quality determining their superiority.}}
As indicated by the rating results, all the videos generated by Narrative Player and those crafted by humans consistently outperformed text. P2 remarked, ``\textit{\visrev{Videos with visuals and animations are way more attractive. They tell the story more vividly and engagingly than plain text.}}'' 
P9 added, ``\textit{\visrev{As the visuals shift, I could tell there is a new insight coming. The animations help me see the links between insights by showing what stayed the same and what changed.}}'' 
However, the text did not invariably score lower than the other four videos. For example, the ``Population'' story in both ablation studies received a lot of ``strong disagreement'' feedback, which scored critically low with mean scores of 2.00 and 2.17. 
P3, while viewing a video from the ablation study on visual generation, noted, ``\textit{With visuals constantly toggling and axes lacking clear relation, the conveyed information seems too muddled. I prefer plain text under this condition.}'' 
\visrev{Similarly, P5 remarked on videos from the ablation study on narrative analysis, ``\textit{The years mentioned in the voiceover seem randomly shown as the whole x-axis or just highlighted points. The x-axis range keeps changing and I can't follow the context.}''}
Ensuring the consistency and contextualization of generated visualization sequences and videos is paramount for users to perceive their superiority over text.

\chirev{\textbf{Automatically generated videos are generally well received by users.}} The user ratings in the study are shown in Fig.~\ref{fig:ratings}. All six narratives scored an average of above 3, with none receiving a ``Strong Disagree'' rating, a unique achievement among the five renditions. ``Population'' \visrev{and ``GDP''} underperformed compared to human-composed videos, whereas the other \visrev{four} narratives matched or surpassed them, notably ``Grades'' (Mean Score: 4.33 vs. 3.08). Against ablation on Narration Analysis, Narrative Player excelled in all but ``GDP'' and ``COVID-19'' narratives, showing no significant advantage in ``GDP'' (Mean Score: 3.33 vs. 3.67) and underperforming in ``COVID-19'' (Mean Score: 3.25 vs. 3.92). For ablation on visual generation, it excelled except for ``Sales'' with similar scores (Mean Score: 3.92 vs. 4.08).
All other ratings are as expected, except for the ``Grades'' of the human-composed version, which will be illustrated in the expert study. 
During experiments, participants occasionally expressed surprise at Narrative Player's capability to capture vague details and their consequent visual representations. P1 commented, ``\textit{The continuously \visrev{rising} line chart in `Population' \visrev{was striking.} Though the text didn't specify the entire range of the x-axis, the algorithm's ability to \visrev{track the timeline's progression was impressive}.}'' While reviewing ``Weather'' (Fig.~\ref{fig:teaser}), P4 observed, ``\textit{\visrev{Presenting temperature and precipitation together was unexpected, yet it clearly illustrated the interplay between sunshine and rainfall.}}'' Such feelings attributed to the analysis of clauses in data fact inference and completion within context. 

\chirev{\textbf{The role of both modules is positively related to narrative complexity.}}
Despite the general lower scores received by videos in both ablation studies, two cases illustrate the impact of narrative complexity on the significance of modules. ``GDP'' and ``COVID-19'' showed no significant score difference between the ablation of narrative analysis and the automatic videos, with both narratives explicitly conveying the data facts. This indicates that when narratives predominantly contain clear facts, or when a narrative's information is sparse but mostly comprises vague clauses with minimal insights, simple replication using clear data facts as anchors is effective. 
For ``Sales'', the absence of significant differences in the visual generation ablation study indicates that if a narrative consistently involves the same data field, resulting in facts with similar \textit{context} and \textit{type}, the impact of the visual generation module's ablation is minimized.

\subsection{Expert \visrev{Study}}
\label{sec: expert study}
\visrev{To provide an in-depth analysis \tvcgminor{of detailed aspects of user engagement} and identify the differences between automatic-generated videos and human-composed ones~\cite{data_player},}
we \visrev{invited} four designers (D1-D4), who daily crafted videos with professional editing tools (e,g., After Effects) \tvcgminor{for more than 3 years}, and three visualization researchers (V1-V3), who have publications in top-tier conferences and journals in VIS and HCI. 
After acquainting them with the background described in Sec.~\ref{sec: preparation} \visrev{and without disclosing the video versions}, they \visrev{watched and rated six sets of narratives, automatic-generated videos, and human-composed ones, used in the user study.} 
\vis{They gave ratings across various dimensions: understandability, memorability, focused attention, enjoyment, consistency, and contextualization. 
The first four dimensions aim to boost user experience, with a focus on the last two due to their technical importance and the challenge of quantifying them without ground truth. All three versions were rated based on the first four dimensions, but only two videos were assessed on the last two, given their visual relevance.}
\visrev{After the ratings, we revealed the video versions and conducted interviews where experts explained their ratings and analyzed the differences between automatic and manual production.}

\begin{figure}[t]
  \centering
  \includegraphics[width=\linewidth]{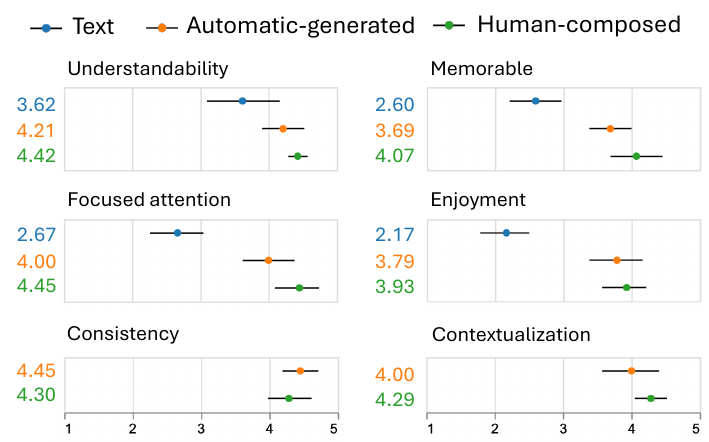}
  \caption{\vis{Expert ratings on six dimensions for text and two kinds of videos with different color legends. Means are represented by shapes, while their 95\% CIs are illustrated with error bars.} } 
  \label{fig:expert ratings}
  \vspace{-10pt}
\end{figure}

Overall, \visrev{as shown in Fig.~\ref{fig:expert ratings}, both videos significantly outperform the text across all dimensions, with the human-composed ones being rated marginally higher than the automatic-generated ones.} All experts concurred that visual sequences generated by Narrative Player align reasonably well with the text and effectively capture and showcase contextual relationships.
Thereby, it mitigates the limitations imposed by gaps in textual information\vis{, as evidenced by the high understandability rating.} These sequences liberate viewers from the confines of information-dense text, enabling a more vivid understanding of the data story and providing new presentations\vis{, as highlighted by significant differences in the three dimensions for engagement.
They praised the alignment between text and visuals in the generated videos and the consistency across visuals, with ratings on par with those of human-composed videos for the two technical dimensions.
}
\visrev{Under the same design space, human-composed versions take several hours to iterate, while Narrative Player generates comparable videos in minutes.}

\visrev{In our interviews, experts shared their reasons for ratings and analyzed differences between automatic and manual versions.} The distinction was most evident in the utilization of annotations. D1 noted, ``\textit{Human-created videos typically featured \visrev{detailed} annotations, such as the bolder red line and clearer text labels in `Weather', which were \visrev{missing} in the automatically generated videos.}'' In fact, automating decisions on when to annotate, and determining its optimal size and position—considering potential occlusions and placements of visual elements—is inherently challenging. Moreover, human-created videos frequently incorporated refined treatments for the introduction of video. For instance, in cases like ``Population'' and ``GDP'', they began with a complete visualization without a visual focus to set the background and subsequently transitioned or highlighted specific data points. Such approaches were not uniformly adopted across cases, which depended on whether the introductory sentence had a clear link with the subsequent content. Participants found such flexible treatment engaging, whereas Narrative Player's consistent choice to begin with a blank visualization appeared somewhat formulaic.

In addition, experts believe that we need to \visrev{further investigate} the design space of visualization \visrev{and user preference} in the video.
The first issue concerns the control of visual elements: V1 particularly appreciated the absence of a legend at the beginning of the human-created ``Sales'', which was introduced only when the data dimensions escalated, while Narrative Player currently didn't support it. ``\textit{Gradually introducing visual elements by the textual content helps me focus on the key points,}'' V1 commented. It suggests that the elements of text-visual mapping in narrative visualization need to be fine-tuned at the level of individual visual elements. 
The second issue concerns varying comprehension of visualization across different backgrounds. D2, like some participants in the user study, felt that the ``Grades'' of the human-composed version were not satisfactory because it was hard to quickly understand insight (the change in grades) from a scatter plot using two test scores as the x and y axes, and the highlight in a single chart is a bit boring. In contrast, V1 and V2 disagreed, noting that visual researchers often have a stronger grasp of such visuals, while designers may favor more engaging animations. This highlights the preferences of different audiences.


Moreover, D1 noted that visuals sometimes should convey information not explicitly stated in the narration. We partially achieved this through Narrative Player's data fact inference and completion within context. Besides, this requirement involves employing visual patterns that transcend local text to describe the global narrative structure, which Narrative Player does not support. For instance, D1 highly praised the human-composed version of ``Population'', as it concurrently maintains both the line chart for population data and the bar chart for changing rates even if the latter part of the narrative doesn't contain information about percentage change. \visrev{Correspondingly,} D1  critiqued the automatic-generated version, which is consistent with the ratings in the user study: ``\textit{These visualizations do adhere to the textual descriptions, but the strict correspondence occasionally disrupts the viewer's thought process by \visrev{sometimes} displaying population growth rates.}'' Similarly, some visuals corresponding to unimportant data facts for vague clauses can be intentionally skipped. Otherwise, although we set the minimal animation duration, frequent transitioning between multiple visualizations and rapidly changing axes may disorient the viewer.

\subsection{\tvcgminor{Example Gallery}}
\tvcgminor{Following existing research that presented example galleries~\cite{data_animator, charticulator} to demonstrate effectiveness and diversity}, we collected diverse datasets and texts from public websites and generated an example gallery with Narrative Player, encompassing various visualization types (e.g., bar, line, point) and narrative themes (e.g., population, weather, COVID-19). Beyond \chirev{Fig.~\ref{fig:teaser} and Fig.~\ref{fig:examples}}, \tvcgminor{more examples can be found \tvcgminorr{on} \url{https://datavideos.github.io/Narrative_Player/}}. 

\section{Discussion}  
\label{sec: discussion}
\chirev{\textbf{Mitigating potential side-effects of LLMs.}} 
Our current system significantly reduces factual inaccuracies through a verification mechanism using sentence embedding and a multi-session approach, ensuring a reliable set of data facts aligns with the narrative. We recognize the potential limitations in the generalizability and accuracy of our method due to the subjective nature of clause classification and parameter setting.
Moreover, challenges such as unstable performance and slower output speed persist. We believe that \chirev{these issues can be addressed by the rapid advancement of LLM research.} 
Interactive tools enabling users to adjust LLM outputs and workflows~\cite{wu2022ai}, such as modifying thresholds or examining clauses, could enhance narrative analysis.

\textbf{Consideration of more complex writing structure \chirev{and longer narrative}.}
Narrative Player currently analyzes paragraphs of the narrative by examining clauses and their neighbors for local context and introduces primary visualization for a global perspective. Yet, long narratives often display structured relationships—parallel, progressive, echoing—that demand a nuanced narrative analysis. 
Narrative structure directly influences visualization selections, with implicit design guidelines suggesting, for instance, consistent visualization styles for parallel or echoing data facts. These patterns have been noticed in the visual storytelling community~\cite{DataParticles2023, Shi2021a}, but how visuals are selected and arranged when the narratives comprise in-depth writing strategy and styles remains to be studied. Moreover, in data videos, insights can be discontinuous and scattered across long segments. For extended narratives, identifying less critical insights or segments lacking data and employing computer vision to create engaging video content is crucial to retain viewer interest \tvcgminorr{in the future}.

\chirev{
\textbf{Exploring more reliable evaluation methods for narrative-driven visual storytelling.} 
The NLP community has begun preliminary studies on evaluating LLMs for understanding and generating coherent stories~\cite{callan2023interesting}, while quantitative evaluation of the technical validity of visual storytelling remains a significant challenge. Due to the diverse preferences and lack of ground truth in storytelling, we currently rely on user studies instead of verifying ``accuracy'' \tvcgminor{of the whole generated videos or individual components. 
Our lab study currently involved a limited number of participants, with each user tasked to watch six sets of materials comprising five versions in random order and provide an overall rating for each version within approximately one hour, while experts provided detailed ratings and analysis for videos. 
This design aimed to ensure both coverage and participant focus and engagement but restricted the scale of the study.
\tvcgminorr{In the future, we plan to conduct more robust and counter-balanced crowdsourcing studies with a Latin-square design involving a larger number of users, which would allow each participant to evaluate only a small subset of large-scale materials with ensured focus.
We could also expand the materials of studies by including more diverse narratives, example videos, and a broad range of human-composed versions from various designers and researchers. 
Additionally, improvements of studies might include designing detailed questionnaires and quizzes about data for the user study, along with metrics to measure the alignment of various components (e.g., facts, visuals, and transitions) between large-scale human-composed versions and automatically generated ones.
}
}
}

\textbf{Advanced considerations beyond consistency and contextualization.} Visuals and videos as educational tools suggest integrating cognitive and educational theories into visual storytelling, especially those related to working memory and cognition. Our use of primary visualization and working memory theory-based modeling in the second module is a step toward incorporating cognitive principles. Indeed, literature on cognitive theory often employs visualization or video as illustrative examples~\cite{schnotz2007reconsideration, castro2019instructional, ayres2007making} and studies in infovis~\cite{eeg4cognitive, cognition4infovis} have integrated cognitive theory to examine the efficacy and complexity of individual visualizations. Extending such approaches to animated visualization sequences and videos could prove beneficial.  Additionally, addressing personal and emotional expression, as highlighted by user feedback, is crucial. Narrative Player lacks automated emphasis detection in narratives and adjustable visual and audio cues to highlight key points~\cite{xie2023creating, li2023geocamera, lan2021kineticharts}. To meet user needs for customization and emotional expression, future enhancements could include more design options and interactive settings for users to personalize their experience, alongside employing contextualized icons or glyphs for more engaging videos.

\chirev{
\textbf{Extension of our pipeline.} The narrative analysis module's handling of ambiguous semantics provides a reference for generating contextualized data facts based on long narratives, and it can be expanded to generate or assist in creating data comics, scrollytelling, and other narrative visualizations related to data facts~\cite{wang2019datashot, lu2021automatic}. Similarly, the consideration of multiple visualizations, visual focus, and primary visualizations in the visual generation module can be applied to other works for sequencing and ordering. Our pipeline could also be generalized to broader authoring with visuals as input: imagine a data analyst creating a data video by initially conducting simple data analysis and writing a data narrative, with the process-generated visuals serving as additional visual input, which could be used as fixing visuals for certain key clauses or primary visualizations in the sequence. In addition, there are also some scenarios where authors do not create visuals. For example, using LLM to interpret data tables and seamlessly craft narratives \tvcgminorr{could be} a viable creative approach. The Narrative Player, as it stands, \tvcgminorr{would be} effectively suited for those creators without requiring modifications.
}

\textbf{Limitations and future work.}
\tvcgminor{Currently, it is an end-to-end automated tool, and} we aim to expand it into a full-scale authoring platform for visual storytelling. 
\tvcgminor{The platform should} autonomously generate and refine narratives based on data input enabled by human-LLM co-writing experience, with options for human-in-the-loop customization~\cite{li2023survey,ren_re-understanding_2023}.
\chirev{For example, users may directly specify color palettes, the primary visualization, and determine the visuals corresponding to certain clauses as anchors.} 
This streamlines the creation of personalized, high-quality data videos while reducing computational load. 
Additionally, narration may mention complex data with high dimensions \visrev{or incorporate multiple facts with varied insights.}
\visrev{To manage this complexity, two approaches could be considered: selecting the facts with the most important insight to avoid overly complex visuals, and enhancing visual expressiveness to cover more facts and comprehensive semantics.}
\visrev{Moreover, the visual types and transitions available in Narrative Player are limited to those supported by GraphScape~\cite{Kim2017b} and Gemini~\cite{kim2020gemini}.} 
We will explore incorporating a broader variety of visualizations and associated visual cues to \visrev{enhance visual expressiveness and} tell more diverse stories. 
\visrev{Also, the default Vega-Lite settings currently used to map facts to visualizations can result in issues like narrow layouts or small visual marks. 
In the future, we need to consider more aesthetic and functional factors in visual design, such as responsiveness for layouts and visual elements.}

\section{Conclusion}

This paper introduces Narrative Player, a tool that generates animated visualization sequences from data narratives and tables, emphasizing consistency and contextualization. Leveraging LLMs and sentence embedding, it interprets narratives to extract contextualized data facts. After mapping facts to visualizations, it optimizes them into a coherent sequence enriched with transition animations. The resultant data video, combining animated visuals and audio narration, received positive feedback in the user study and expert study, affirming its comparability to human-created videos and the effectiveness of its core modules. Encouraged by the feedback, we aim to further refine Narrative Player to more closely match the quality of real-world data videos.

\section*{Acknowledgments}
The authors want to thank the reviewers for their suggestions. This
work is supported by Natural Science Foundation of China
(NSFC No.62472099 and No.62202105).

\bibliographystyle{abbrv-doi-hyperref}
\bibliography{template}




 



\begin{IEEEbiography}[{\includegraphics[width=1in,height=1.2in,clip,keepaspectratio]{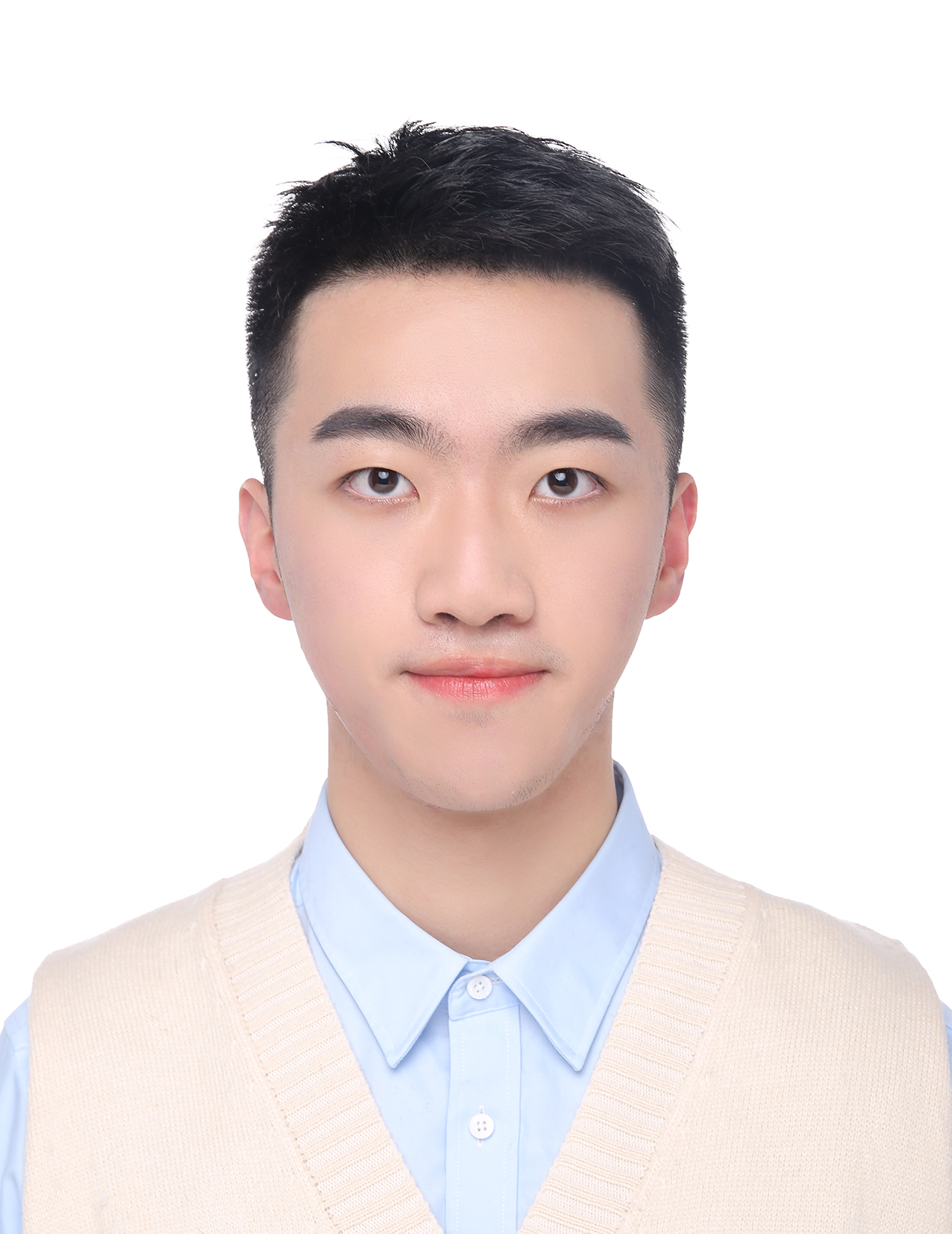}}]{Zekai Shao}
is a PhD student in the School of Data Science at Fudan University. He obtained his bachelor's degree in Intelligent Science and Technology from Fudan University in 2023. His research interests include visual storytelling and evaluation. For more details, please refer to \url{https://zekaishao25.github.io/}. 
\end{IEEEbiography}

\begin{IEEEbiography}[{\includegraphics[width=1in,height=1.2in,clip,keepaspectratio]{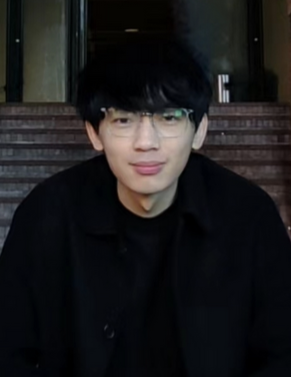}}]{Leixian Shen}
is a PhD candidate in the Department of Computer Science and Engineering at The Hong Kong University of Science and Technology. He received his master's degree in Software Engineering from Tsinghua University in 2023 and obtained his bachelor's degree in Software Engineering from Nanjing University of Posts and Telecommunications in 2020. His research interests include visual data analysis and storytelling. For more details, please refer to \url{https://shenleixian.github.io/}. 
\end{IEEEbiography}

\begin{IEEEbiography}[{\includegraphics[width=1in,height=1.2in,clip,keepaspectratio]{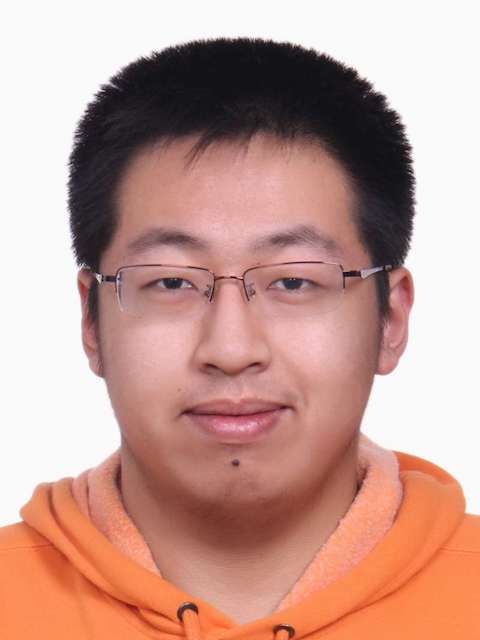}}]{Haotian Li}
is currently a Ph.D. candidate in Computer Science and Engineering at the Hong Kong University of Science and Technology (HKUST). His main research interests are data visualization, visual analytics, and human-computer interaction. He received his BEng in Computer Engineering from HKUST. For more details, please refer to \url{https://haotian-li.com/}. 
\end{IEEEbiography}

\begin{IEEEbiography}[{\includegraphics[width=1in,height=1.2in,clip,keepaspectratio]{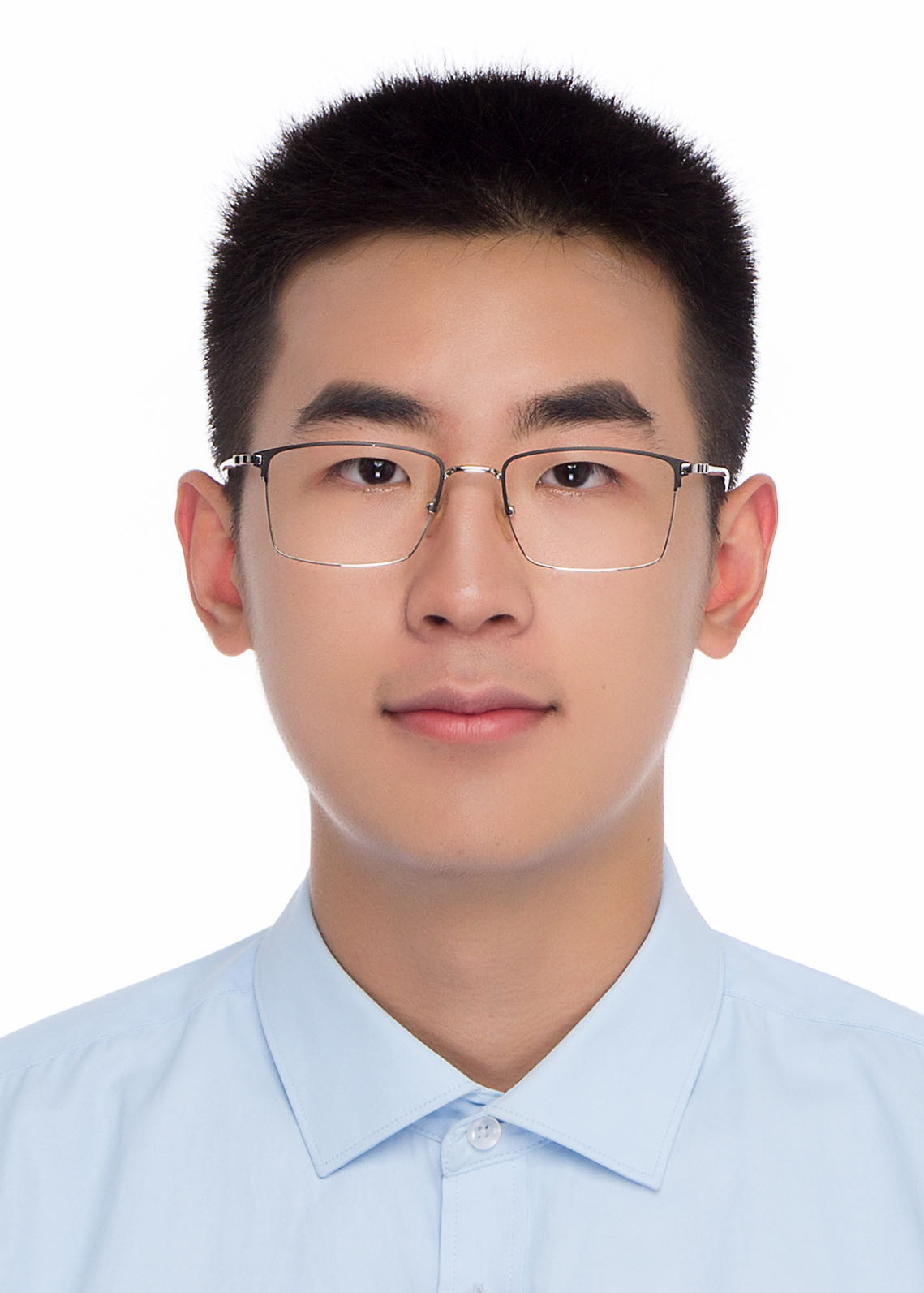}}]{Yi Shan} is a PhD student in School of Data Science at Fudan University. He obtained his bachelor's degree in the school of Computer Science and Technology from Fudan University in 2024. His research interests include visual analytics.

\end{IEEEbiography}

\begin{IEEEbiography}[{\includegraphics[width=1in,height=1.2in,clip,keepaspectratio]{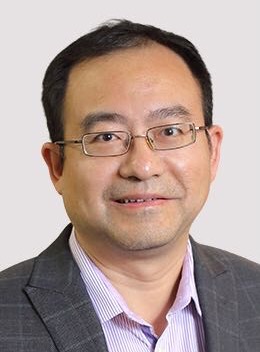}}]{Huamin Qu}
is a chair professor in the Department of Computer Science and Engineering (CSE) at the Hong Kong University of Science and Technology (HKUST) and also the dean of the Academy of Interdisciplinary Studies (AIS) and the head of the Division of Emerging Interdisciplinary Areas (EMIA) of HKUST. He obtained a BS in Mathematics from Xi'an Jiaotong University, China, an MS and a PhD in Computer Science from the Stony Brook University. His main research interests are in visualization and human-computer interaction, with focuses on urban informatics, social network analysis, E-learning, text visualization, and explainable artificial intelligence (XAI). For more information, please visit \url{http://huamin.org/}.
\end{IEEEbiography}

\begin{IEEEbiography}[{\includegraphics[width=1in,height=1.2in,clip,keepaspectratio]{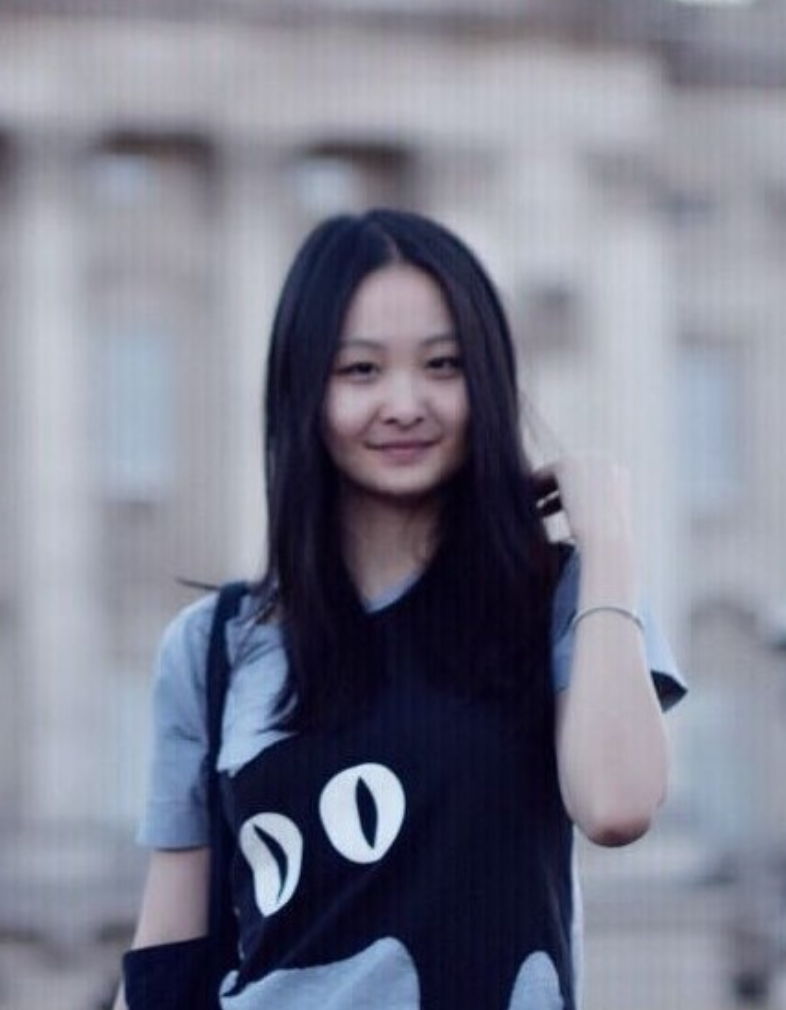}}]{Yun Wang}
is a senior researcher in the Data, Knowledge, Intelligence (DKI) Area at Microsoft. Her research lies in the intersection of Human-Computer Interaction, and Information Visualization. She develops innovative techniques and interactive systems to facilitate Human-AI Collaboration, Human-Data Interaction, Visual Communication, and Data Storytelling through an interdisciplinary approach. She received her Ph.D. from the Hong Kong University of Science and Technology. For more details, please refer to \url{https://www.microsoft.com/en-us/research/people/wangyun/}. 
\end{IEEEbiography}

\begin{IEEEbiography}[{\includegraphics[width=1in,height=1.2in,clip,keepaspectratio]{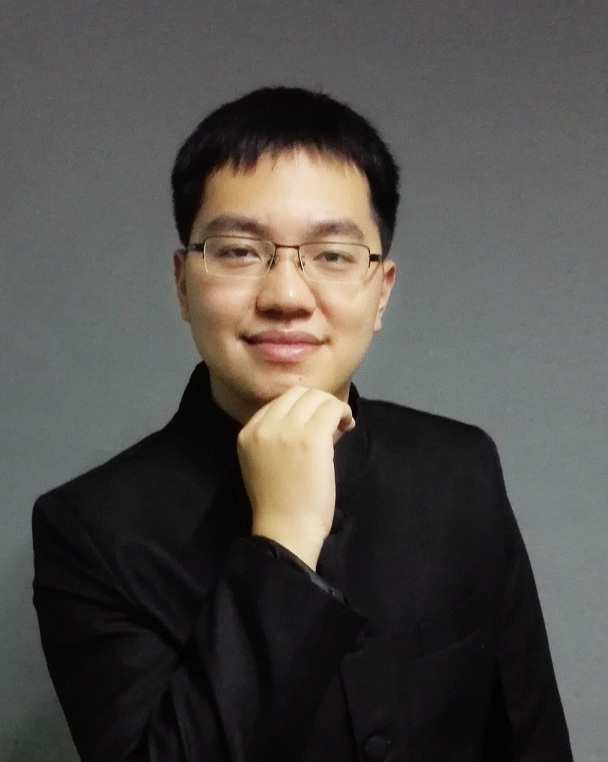}}]{Siming Chen}
is an Associate Professor at School of Data Science, Fudan University. Prior to this, he was a Research Scientist at Fraunhofer Institute IAIS (Intelligent Analysis and Information Systems) and a Postdoc Researcher at the University of Bonn in Germany. He received his Ph.D. in computer science at the School of EECS, Peking University and received his BS degree in computer science at Fudan University. His research interests are visualization and visual analytics, with the emphasis on social media visualization, spatial-temporal visual analytics, and cybersecurity visual analytics. He has published 70 papers and more than 20 in top conferences and journals, including IEEE VIS, IEEE TVCG, EuroVis, etc. He was awarded 10+ best paper/poster awards and honorable mentioned awards in multiple conferences and won multiple IEEE VAST Challenge Excellent Awards. For more information, please visit \url{http://simingchen.me}.
\end{IEEEbiography}


\end{document}